# Two-Dimensional Large Gap Topological Insulators with Large Rashba Spin-Orbit Coupling in Group-IV films

Shou-juan Zhang, Wei-xiao Ji, Chang-wen Zhang,* Ping Li, and Pei-ji Wang

School of Physics and Technology, University of Jinan, Jinan, Shandong, 250022, People's Republic of China

Rashba spin-orbit coupling (SOC) in topological insulators (TIs) has attracted much interest due to its exotic properties closely related to spintronic devices. The coexistence of nontrivial topology and giant Rashba splitting, however, has rare been observed in two-dimensional (2D) films, limiting severely its potential applications at room temperature. Here, we through first-principles calculations to propose a series of inversion-asymmetric group-IV films, $ABZ_2$ ($A \neq B$ = Si, Ge, Sn, Pb; $Z$ = F, Cl, Br), whose stability are confirmed by phonon spectrum calculations. The analyses of electronic structures reveal that they are intrinsic 2D TIs with a bulk gap as large as 0.74 eV, except for $GeSiF_2$, $SnSiCl_2$, $GeSiCl_2$ and $GeSiBr_2$ monolayers which can transform from normal to topological phases under appropriate tensile strain of 4, 4, 5, and 4 %, respectively. The nontrivial topology is identified by $Z_2$ topological invariant together with helical edge states, as well as the berry curvature of these systems. Another prominent intriguing feature is the giant Rashba spin splitting with a magnitude reaching 0.15 eV, the largest value reported in 2D films so far. The tunability of Rashba SOC and band topology can be realized through achievable compressive/tensile strains (-4 ~ 6%). Also, the BaTe semiconductor is an ideal substrate for growing $ABZ_2$ films without destroying their nontrivial topology. These findings present a platform to explore 2D inversion-asymmetric TIs for room-temperature device applications.

Corresponding author: ss_zhangchw@ujn.edu.cn

# I. Introduction

Topological insulators (TIs) are new states of quantum matter interesting for fundamental condensed matter physics and important for practical applications [1-3]. In particular, for two-dimensional (2D) TIs, all the scatterings of conducting edge states are completely forbidden, while the surface states of three-dimensional (3D) TIs are not protected against scattering at any angles other than $180^\circ$. The helical edge states of nanoribbons made from 2D TIs provide massless relativistic carriers with intrinsic spin-momentum lock and negligible backscattering, ideal for the realization of conducting channels without dissipation in electronic devices [4, 5]. However, quantum spin Hall (QSH) effect has only been experimentally observed in HgTe/CdTe [6, 7] and InAs/GaSb [8, 9] quantum wells at very controlled and low temperature. In spite of this successful progress, the experimental realizations of these topological quantum states in 2D films are still challengeable.

Graphene,[1, 2] the first 2D TI model, supports an unobservable small bulk gap due to its extremely weak spin-orbit coupling (SOC). Afterwards, several candidate materials are proposed to overcome the issue in graphene for the observation of QSH effect. These samples include silicene, [10] germanene[11], stanene[12], plumbene,[13] bismuth[14], and arsenene[15]，in which they exhibit the topology properties alike to grapheme, but their nontrivial bulk gaps are still small. Motivated by the experimental synthesis of hydrogenated germanene (GeH) [16], the chemical functionalization on honeycomb structure, e.g., Bi/Sb/As [17-20] and Ge/Sn/Pb films [21-27], has been proposed to host QSH effect with enhanced band gaps. For example, Xu *et* [12] reported that the nontrivial band gap of pristine stanene is merely 0.1eV, while it is enlarged to 0.3 eV under fully halogenation [12, 23-26]. More interestingly, the bulk gaps of 2D plumbene films with chemical decoration, PbX (X= H, F, Cl, Br and I) monolayers, have been elevated to a remarkable 1.34 eV [13] the largest gap of all the reported 2D TIs as we know. All these results provide a great potential for the realization of QSH effect at room temperature in device paradigms for quantum information processing.

Rashba SOC, on the other hand, generally leads to spin-polarized band dispersion curves with in-plane opposite helical spin texture [28]. This feature originates from structural inversion-symmetry breaking at the surface/interface system, allowing the control of spin direction through an electric field [29]. As compared with inversion-symmetric TIs, these systems are called 2D inversion-asymmetric TIs (IASTIs), which are more promising due to their perfect performance in realizing new topological phenomena, such as crystalline-surface dependent topological electronic states [30], pyroelectricity [31], natural topological *p-n* junctions [32], and so on. In conventional semiconductor materials [33-36], Rashba SOC can only be tuned in a relatively limited range under electric field since it is generally rather weak, so it would be one of the important reasons to interpret the great attention the Datta-Das spin FET [37], which still has not been realized in experiments. Recently, Arguilla *et al* [38] synthesize 2D honeycomb $Ge_{1-x}Sn_xH_{1-x}(OH)_x$ film from the topochemical deintercalation of $CaGe_{2-2x}Sn_{2x}$. In comparison with inversion-symmetric systems [15-31], the introduction of tin breaks the inversion-symmetry of 2D germanene, which may induce a significant large Rashba SOC. Motived by above suppose, we attempt to extend and advance the group-IV 2D films to realize giant Rashba SOC accompanied with nontrivial topology, which may our proposal achievable experimentally accessible.

In present work, we report on the electronic and topological properties of a series of 2D IASTIs in group-IV binary films, which are called as $ABZ_2$ monolayers, where *A* and *B* belong to group-IV elements (Si, Ge, Sn, Pb), as well as *Z* is the decorated halogen atoms (F, Cl, and Br), as illustrated in Fig. 1 (a). We find that most of these group-IV binary films are demonstrated as 2D IASTIs possess sizeable band gaps, reaching a maximum value of 0.74 eV for $PbSnCl_2$, which exceed the thermal energy at room temperature. Though $GeSiF_2$, $SnSiCl_2$, $GeSiCl_2$ and $GeSiBr_2$ monolayers are normal insulators (NI), they can transform 2D TI phases under tensile strain of 4, 4, 5, 4 %, respectively. The nontrivial quantum states are identified by $Z_2$ topological invariant together with helical edge states induced by SOC, as well as berry curvature of these systems. The most prominent feature in these systems is the giant Rashba splitting energy, for example, a value of 0.15 eV for PbSiCl$_2$, which is the largest value reported in 2D films so far. Furthermore, a favorable tunability of Rashba SOC and nontrivial topology can be obtained through achievable compressive/tensile strains (-4 ~ 6%). We also find that the Te(111)-terminated surface in semiconductor BaTe is an ideal substrate for experimental realization of these monolayers, without destroying their nontrivial topology. The coexistence of giant Rashba SOC and band topology in these films may bring rich spin-related phenomena in 2D IASTIs, which offers a special contribution to nanoelectronics and spintronics.

## II. Calculated methods and details

All calculations are carried out using the plane wave basis Vienna ab initio simulation pack (VASP) code [39,40] implementing density functional theory (DFT). The projector-augmented wave (PAW) method [41] is used to describe electron-ion potential. The exchange-correlation potential is approximated by generalized gradient approximation (GGA) in the Perdew-Burke-Ernzerhof (PBE) [42] form. Considering the possible underestimation of GGA method, the Heyd-Scuseria-Ernzerhof (HSE06) hybrid functional is employed to check the band topology [43]. We use an energy cutoff of 450 eV and maximum residual force less than 0.001 eV/Å. Periodic boundary conditions are employed to simulate these 2D systems, and the Brillouin zone is sampled by using a 17×17×1 Gamma-centered Monkhorst-Pack grid. Moreover, SOC is taken into account in terms of the second-variational procedure. The phonon dispersion calculations are carried out by employing the PHONOPY code [44] through the density-functional perturbation theory (DFPT) approach.

## III. Results and discussion

Fig. 1(a) displays the honeycomb lattice of $AB$Cl$_2$ ($A \neq B$ = Si, Ge, Sn, Pb) monolayers, accompanied with its Brillouin zone in Fig. 1(b). All these configurations are iso-structural to inversion-symmetric Ge/Sn/Pb monolayers with C$_{3v}$ symmetry [19-25], but the *A* and *B* sublattices are occupied with different group-IV atoms, resulting in inversion-symmetry breaking. To verify the geometric stability of these structures, the total energies $AB$Cl$_2$ monolayers with respect to lattice constant *a* are illustrated in Fig. 1(c), where *a* is varied to identify structural ground state. Interestingly, the double-well energy variation curves occur in all cases, namely, the high-buckled (HB) and low-buckled (LB) states, respectively. In HB structure, the vertical layer distance lies in the range of 3.58 − 5.15 Å, while the distance in LB state is smaller within 1.42−2.51 Å, as listed in Table I. The LB structure here, however, is more stable by at least 1.04 eV per unit cell than HB structure, different form the cases of III-Bi films, where the HB structure is the ground state. Thus, we only focus on stable LB chlorinated monolayers in the following. Furthermore, the energetically structural stability is evaluated by the formation energy expressed as

$$\Delta E_f = E(ABCl_2) - E(AB) - 2\mu(Cl)$$

where $E(ABCl_2)$ and $E(AB)$ are the total energies of chlorinated and pristine AB monolayers, respectively, while $\mu(Cl)$ is the chemical potential of chlorination atoms. As listed in Table I, the calculated formation energies are in the range of -2.284 ~ -3.146 eV per atom, indicating no phase separation emerges in these systems. For practical application as a nanodevice, the 2D films should be chemically inert and environmentally friendly. Thus, the phonon spectrum dispersions are calculated and plotted in Fig. 2, which demonstrates that all $AB$Cl$_2$ monolayers are dynamically stable since all the vibrational models are real in the whole Brillouin zone.

The band structures of $AB$Cl$_2$ monolayers are shown in Fig. 3, with the orbital-projected components displayed in Fig. S1 in the supplementary information. In the absence of SOC, the PbSnCl$_2$, PbGeCl$_2$ PbSiCl$_2$ and SnGeCl$_2$ monolayers show gapless semiconductor feature with valence band maximum (VBM) and conduction band minimum (CBM) degenerated at the Γ point. The Fermi level locates exactly at the degenerated point with a zero density of states. Notably, the slopes of the valence band and conduction band in such a touching point show the parabolic dispersion, instead of a linear one, as illustrated by a 3D band structure for PbSnCl$_2$ monolayer (Fig. 4 (a)). This scenario for the electronic properties is quite similar to Ge/Sn/Pb monolayers [21-27]. Upon including SOC, the band structures produce a transition from semimetal to semiconductor, as the degenerated states at the touching point separate from each other, see Fig. 3(a)-(d) and Fig. 4(b). As observed in previously reported Bi/Sb/As [17–20], the SOC-induced band-gap opening at the Fermi level is a strong indication of the existence of topologically nontrivial phases. Furthermore, from Si to Pb atom, the band gap opening gradually increases (Table I), since the SOC effect gets stronger due to increasing atomic numbers of heavier atoms. Here we label this SOC-induced band gap at the $\Gamma$ point as $E_{\Gamma\text{-soc}}$, to be distinguished from whole gap ($E_{g\text{-soc}}$), as listed in Table I. To eliminate the possible underestimation of the band gap, we employ additional HSE06 [37] to confirm the existence of band inversion for all these systems. As expected, the values of $E_\Gamma$ are enhanced to 0.74, 0.49, 0.41 and 0.23 eV, respectively, for $PbSnCl_2$, $PbGeCl_2$, $PbSiCl_2$ and $SnGeCl_2$ monolayers.

To confirm the nontrivial topological nature, we introduce the evolution of Wannier Center of Charges (WCCs) [45] to calculate $Z_2$ topological invariant due to inversion-symmetry breaking. The Wannier functions (WFs) related with lattice vector R can be written as:

$$|R,n\rangle = \frac{1}{2\pi}\int_{-\pi}^{\pi} dk e^{-ik(R-x)}|u_{nk}\rangle$$

Here, a WCCs $\bar{x}_n$ can be defined as the mean value of $\langle 0n|\hat{X}|0n\rangle$, where the $\hat{X}$ is the position operator and $|0n\rangle$ is the state corresponding to a WF in the cell with $R = 0$. Then we can obtained

$$\bar{x}_n = \frac{i}{2\pi}\int_{-\pi}^{\pi} dk \langle u_{nk}|\partial_k|u_{nk}\rangle$$

Assuming that $\sum_\alpha \bar{x}_\alpha^S = \frac{1}{2\pi}\int_{BZ} A^S$ with $S = I$ or $II$, where summation in $\alpha$ represents the occupied states and $A$ is the Berry connection. So we have the format of $Z_2$ invariant:

$$Z_2 = \sum_\alpha [\bar{x}_\alpha^I(T/2) - \bar{x}_\alpha^{II}(T/2)] - \sum_\alpha [\bar{x}_\alpha^I(0) - \bar{x}_\alpha^{II}(0)]$$

The $Z_2$ invariant can be obtained by counting the even or odd number of crossings of any arbitrary horizontal reference line. Taking PbSnCl$_2$ monolayer as an example, we present the evolution lines of WCCs in Fig. 5(a). One can see that the WCCs evolution curves cross any arbitrary reference lines odd times, yielding $Z_2 = 1$.

The most important character of 2D TIs is the appearance of helical edge states with spin-polarization protected by TRS. Here we compute the edge Green's functions using the *ab initio* TB model. The *ab initio* TB model is constructed by downfolding the bulk bands obtained by DFT calculations, using the recursive method in terms of maximally localized Wannier functions (MLWFs) [46]. Figures 4(c) and 4(d) show the orbital-decomposed band structures, in which the red and blue circles represent group-IV *s* and $p_{x,y}$ states, respectively, and the symbol size is proportional to the orbital weight. Obviously, the bulk energy bands near the Fermi level are predominantly from $p_{x,y}$ orbitals, thus the MLWFs are derived from atomic *p*-like orbitals. In this case, the TB bulk bands perfectly reproduce the DFT bands around the Fermi level up to ± 1.0 eV, as illustrated in Fig. 5(b). Fig. 5(c) displays the local density of states (LDOS) of the calculated edge states. Indeed, all the edge bands for PbSnCl$_2$ monolayer connect completely the conduction and valence bands and span 2D bulk gap, yielding a 1D gapless edge states, in good agreement with the $Z_2$ invariant. By identifying spin-up and spin-down contributions in edge spectral function, the counter-propagating edge states exhibit opposite spin-polarization (Fig. 5(d)), which benefit from their robustness against nonmagnetic scattering.

For GeSiCl$_2$ and SnSiCl$_2$ monolayers, on the other hand, they exhibit semiconductor feature with direct band gap of 0.285 and 0.265 eV in the absence of SOC, as shown in Figs. 3(e) and 3(f). As we know that external strain is an effective avenue to tune the band gap, this may introduce a band inversion by SOC. To simulate strain effects on the electronic structures, we employ the external strain to these systems which is defined as $\varepsilon = (a-a_0)/a_0$, where *a* and $a_0$ are lattice constants with and without strains, respectively. In Fig. 6 we present the band structures with respect to external strains of GeSiCl$_2$ as an example, where the red and blue colors represent *s* and $p_{x,y}$ states, respectively. One can see that the band structure of GeSiCl$_2$ is sensitive to the lattice expansion. Without SOC, the CBM ($s$) and VBM ($p_{x,y}$) move toward each other, leading to a reduction of band gap. The band gap closes at the $\Gamma$ point when $\varepsilon = 5\%$, as illustrated in Fig. 6(c), suggesting semi-metallic character at the Fermi level. When the tensile strains are beyond 5%, the $p_{x,y}$ orbitals shifted above $s$ orbital, see Fig. 6(d). As a consequence, the opposite spatial shift of VBM and CBM decreases the band gap, which is essential to topological phase transitions. With increasing biaxial strain, the energy difference ($E_{s\text{-}p}$) between the $s$ and $p_{x,y}$ orbitals at the $\Gamma$ point near the Fermi level increase monotonically, but the system remains the semi-metallic feature. When SOC is included, a sizable band gap opens, see Figs. 6(e)-(h). Similar results are also found in $SnSiCl_2$ monolayer in Fig. 7, but its critical point of elastic strain is 4 %, smaller than $GeSiCl_2$ monolayer. Such an SOC-induced band gap reopening usually corresponds to a NI - TI transition, as illustrated in Fig. 8.

To gain a physical understanding of QSH effect, we analyze the orbital contribution around the Fermi level for $GeSiCl_2$ monolayer. Figure 9 presents the band evolution at the $\Gamma$ point, in which the energy levels near the Fermi level are mainly composed of $s$ and $p_{x,y}$ orbitals from group-IV elements. The chemical bonding between Si-Ge atoms makes it split into the bonding and antibonding states, *i.e.*, $s^{\pm}$ and $p_{x,y}^{\pm}$, respectively. At ground state in Fig. 9(a), the bands near the Fermi level are contributed by the $p_{x,y}^{+}$ and $s^{-}$, with the $s^{-}$ being above $p_{x,y}^{+}$, indicating a normal band order. When considering the external tensile strains, as illustrated in Fig. 9(b), the enlarged lattice constant weakens the interactions between A and B sublattices, decreasing the splitting between the bonding and antibonding states, which lowers $s^{-}$ and raises $p_{x,y}^{+}$ level accordingly. Thus, depending on the strength of interatomic coupling, the band gap of $GeSiCl_2$ can be continuously reduced, with the band order being reversed at critical point ($\varepsilon = 5\%$), making it a 2D TI with $Z_2 = 1$. In another words, the level crossing leads to a parity exchange between occupied and unoccupied bands, inducing a NI-TI phase transition. While for $PbSnCl_2$, $PbGeCl_2$ $PbSiCl_2$ and $SnGeCl_2$ monolayers, the lattice constant (See Fig. 1(c) and Table I) is enlarged significantly at equilibrium state, equal to strain-induced $GeSiCl_2$, thus they are native 2D TIs without any strain. Herein, we must point out that, although the $s$-$p_{x,y}$ band inversion is caused mainly by the strength of interatomic coupling, the SOC is still indispensable because it makes the $p_{x,y}^+$ orbital split into $p_{x+iy,\uparrow}^+$ & $p_{x-iy,\downarrow}^+$ and $p_{x-iy,\uparrow}^+$ & $p_{x+iy,\downarrow}^+$, opening a larger band gap. In previous works, the topological feature mainly originates from $p_z$ states such as graphene and silicene [1-3], but here the band topology is attributed to the contribution of $p_{x,y}$ orbitals on group-IV elements, thus their band gap is enhanced significantly.

Furthermore, to clearly show the QSH effect in these systems and confirm its nontrivial topology, we also calculate the intrinsic spin Hall conductivity of PbSnCl$_2$ monolayer. Using the Kubo formula [47, 48], the spin Hall conductivity tensor is expressed as,

$$\sigma^{SH} = \frac{e^2}{2\hbar} \int \frac{d^2k}{(2\pi)^2} \Omega^{SH}(k)$$

Here the spin Berry curvature is given by

$$\Omega^{SH}(k) = -2 \sum_{E_{n,k} \leq E_F} \sum_{n' \neq n} \frac{Im < \Psi_{n,k}|j_x|\Psi_{n',k} >< \Psi_{n',k}|v_x|\Psi_{n,k} >}{(E_{n,k} - E_{n',k})^2}$$

where $n$ is the band index, $\psi_{n,k}$ is the eigenstate, $v_y$ is the velocity operator, and $j_x$ is the spin current operator defined as $(s_z v_x + v_x s_z)/2$. Also, a factor of $2e/\hbar$ in $\sigma^{SH}$ is multiplied to convert its unit into the conventional electron conductivity $e^2/h$. Figure 10(a) displays the $k$-resolved spin Berry curvature at the Fermi level. One can clearly see the positive peak of $\Omega^{SH}(k)$ around the Γ point, which is consistent with its band topology. The spin Hall conductance *via* its chemical potential shows a nearly quantized value $2e^2/h$ of conductance within the energy window of the SOC gap ranging from about − 0.2 eV to the Fermi energy, demonstrating that the QSH effect could be easily observed experimentally at room temperature.

The most prominent feature in inversion-asymmetric systems is the existence of giant Rashba SOC in valence band edge rather than Dresselhaus splitting [49]. These can be confirmed by the spin texture of Rashba spin splitting in Fig. 10(c), in which the spins of the electronic states $\varepsilon_\pm(k)$ are oppositely aligned within the $k_x$ and $k_y$ plane, and are normal to the wave vector **k**. Here, we magnify the Rashba splitting energy $E_R$ and the momentum offset $k_R$ in Fig. 10(b), which corresponds to the splitting of the VBM along momentum direction at the Γ point, indicated by the rectangular box in Fig. 3(c). According to the linear Rashba model of the 2D systems, the dispersion of Rashba splitting can be expressed as $\varepsilon_\pm(k) = \hbar^2 k^2 / 2m^* \pm \alpha_R |k|$ with k = ($k_x$, $k_y$), where the $m^*$ is effective mass and $\alpha_R = e\eta_{so}E$ is the Rashba constant. Thus we can obtain a giant Rashba parameter eV/Å than the well-known oxide interface LaAlO$_3$/SrTiO$_3$ (0.01–0.05 eV/Å) [50-52], quantum well InGaAs/InAlAs (0.07eV/Å)[53]and Au(111) surface [54]. These great Rashba splitting energy $E_R$ and momentum offset $k_R$ are desired for stabilizing spin and achieving a significant phase offset for different spin channels.

It is noteworthy that strain engineering can not only tune the topological phase, but also change the Rashba spin splitting significantly in these monolayers. In Fig. 11, we present three parameters ($E_R$, $k_R$, $\alpha_R$) under different biaxial strains by solid triangles for PbSiCl$_2$ as an example. These three parameters decrease monotonically with the increasing lattice constants, demonstrating that a compressive/tensile strain can enhance/decrease the Rashba SOC strength. For example, the Rashba splitting can reach up to 0.182 eV under -4 % strain. Thus, by biaxial strain engineering, we can enhance or reduce the SOC strength according to the experimental demands. The Rashba spin splitting of energy bands obtained here provides a chance for spintronic device applications without magnetic field, for instance, spin field-effect transistor (FET).

Finally, we must point out that that the chlorination in group-IV monolayers is not the only way to achieve QSH effect, the same results can be obtained by decorating the surface with other halogens, such as F and Br atoms. As evidenced in Fig. S2 in the supplementary information, the double-well energy curve still exists, with the LB structure being a ground state. The corresponding band structures of ABF$_2$ and ABBr$_2$ monolayers with and without SOC are further plotted in Figs. 12 and 13. Most of halogenated ABX$_2$ monolayers are similar to chlorinated ones with a topological invariant $Z_2 = 1$, except for GeSiF$_2$ and GeSiBr$_2$ with $Z_2 = 0$. However, they can transform into QSH insulators under external strain of 4 %, as illustrated in Figs. 8(a) and (c). In addition, Figure 8 presents the calculated band gaps at equilibrium states, accompanied with the lattice parameters listed in Table SI and SII. Interestingly, the nontrivial band gaps can reach as large as 0.11 - 0.63 eV (Fig. 10(d)), which is comparable or larger than stanene (0.1 eV) [12] or bismuth (0.2 eV) [14]. Such large nontrivial band gaps are capable of stabilizing the boundary current against the influence of thermally activated bulk carriers, and thus are beneficial for high-temperature applications.

Experimentally, choosing a suitable substrate is a key factor in device application, since a free-standing film must be eventually deposited or grown on a substrate. Considering the small lattice constant mismatch between $PbSnCl_2$ and semiconductor BaTe [55], we select Te-terminated (111) surface of semiconductor BaTe to support $PbSnCl_2$ monolayer, as illustrated in Fig. 14. In the equilibrium state, the bottom (top) Sn atoms in $PbSnCl_2$ are located at top (hcp) sites of the substrate, while the top Pb atoms are located at hcp sites, see Figs. 14(a) and (b). On this Te-terminated surface, Sn atoms bind preferably on top of Te atoms by forming chemical bonds, so the $p_z$ orbital of group-IV atoms and dangling bond of the substrate are both fully passivated. Figures 14(c)-(f) present the band structures with and without SOC, respectively. One can see that a few bands appear within the bulk gap of substrate around the Fermi level, which is mostly contributed by group-IV atoms according to wave-function analysis. By projecting Bloch wave functions onto atomic orbitals of $PnSnCl_2$, Bloch states contributed by the s orbital of group-IV atoms are visualize by red dots. These states shift to valence bands around the Γ point, suggesting a clear *s-p* band order. This SOC-induced band gap is topologically nontrivial as explained by the band inversion, thus it is a robust 2D TI on Te-terminated substrate of BaTe semiconductor.

### IV. Conclusions

In summary, by means of first-principles calculations, we predict a series of inversion-asymmetric group-IV monolayers, $ABZ_2$ ($A{\neq}B$ = Si, Ge, Sn, Pb; $Z$ = F, Cl,

Br), allows for the simultaneous presence of topological order and Rashba SOC. The phonon spectrum calculations reveal that they are dynamically stable. The analyses of electronic structures reveal that most of these monolayers are native 2D TIs with a bulk gap as large as 0.74 eV, while GeSiF$_2$, SnSiCl$_2$, GeSiCl$_2$ and GeSiBr$_2$ can transform from normal to topological phases under tensile strain of 4, 4, 5 and 4 %, respectively. The nontrivial topology is identified by $Z_2$ topological invariant together with helical edge states, as well as the berry curvature of these systems. The most intriguing feature is their giant Rashba spin splitting with a magnitude reaching 0.15 eV for PbSiCl$_2$, the largest value reported in 2D films so far. A large tunability of Rashba SOC and band topology can be realized through achievable compressive/tensile strains (-4 ~ 6%). In addition, the Te(111)-terminated surface in semiconductor BaTe is predicted to be an ideal substrate for experimental realization of these systems, without destroying their nontrivial topology. Considering their simple crystal structure as well as large nontrivial band gap and giant Rashba splitting energy in these monolayers, these findings present a platform to explore 2D inversion-asymmetric TIs for room-temperature device applications.

**Acknowledgements**

This work was supported by National Natural Science Foundation of China (Grant Nos. 11434006 and 11304121).

**Table I**. Calculated equilibrium lattice parameters $a$(Å), the buckled height $h$(Å), band Gaps $E_g$(eV), band Gaps with SOC $E_g$ (eV), the band gap located at $\Gamma$ point $E_\Gamma$(eV), gap at $\Gamma$ with SOC $E_{\Gamma\text{-}SOC}$(eV) and topological Invariants ($Z_2$) for six films.

| Structure | $a$(Å) | $h$(Å) | $E_g$(eV) | $E_\Gamma$(eV) | $E_{g\text{-}SOC}$(eV) | $E_{\Gamma\text{-}SOC}$(eV) | $Z_2$ |
|---|---|---|---|---|---|---|---|
| $PbSnCl_2$ | 5.04 | 0.11 | 0 | 0 | 0.720 | 0.739 | 1 |
| $PbGeCl_2$ | 4.93 | 0.55 | 0 | 0 | 0.405 | 0.492 | 1 |
| $PbSiCl_2$ | 4.79 | 0.62 | 0 | 0 | 0.294 | 0.409 | 1 |
| $SnGeCl_2$ | 4.73 | 0.59 | 0 | 0 | 0.175 | 0.234 | 1 |
| $SnSiCl_2$ | 4.48 | 0.71 | 0.344 | 0.344 | 0.265 | 0.265 | 0 |
| $GeSiCl_2$ | 4.24 | 0.66 | 0.334 | 0.334 | 0.285 | 0.285 | 0 |

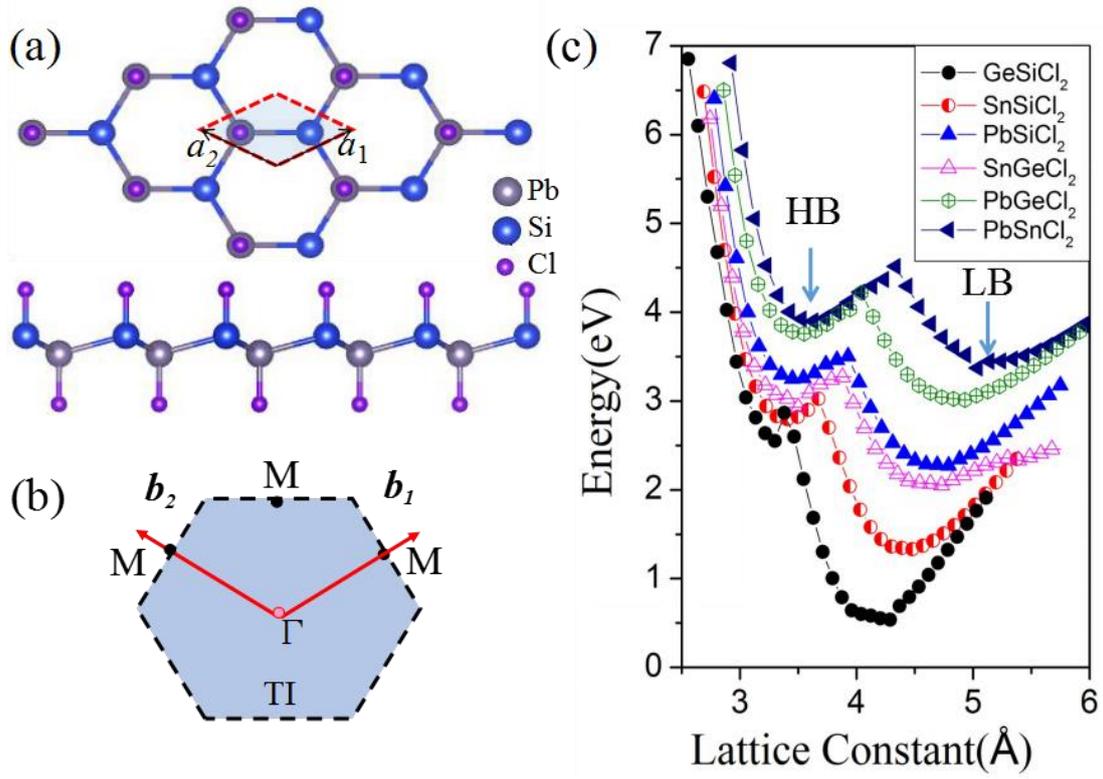

Fig. 1 (a) Top view and side view of 2D PbSiCl$_2$ monolayer. (b) 2D Brillouin zones with specific symmetry points. (c) Total energies of ABCl$_2$ monolayers as a function of lattice constant *a*, respectively.

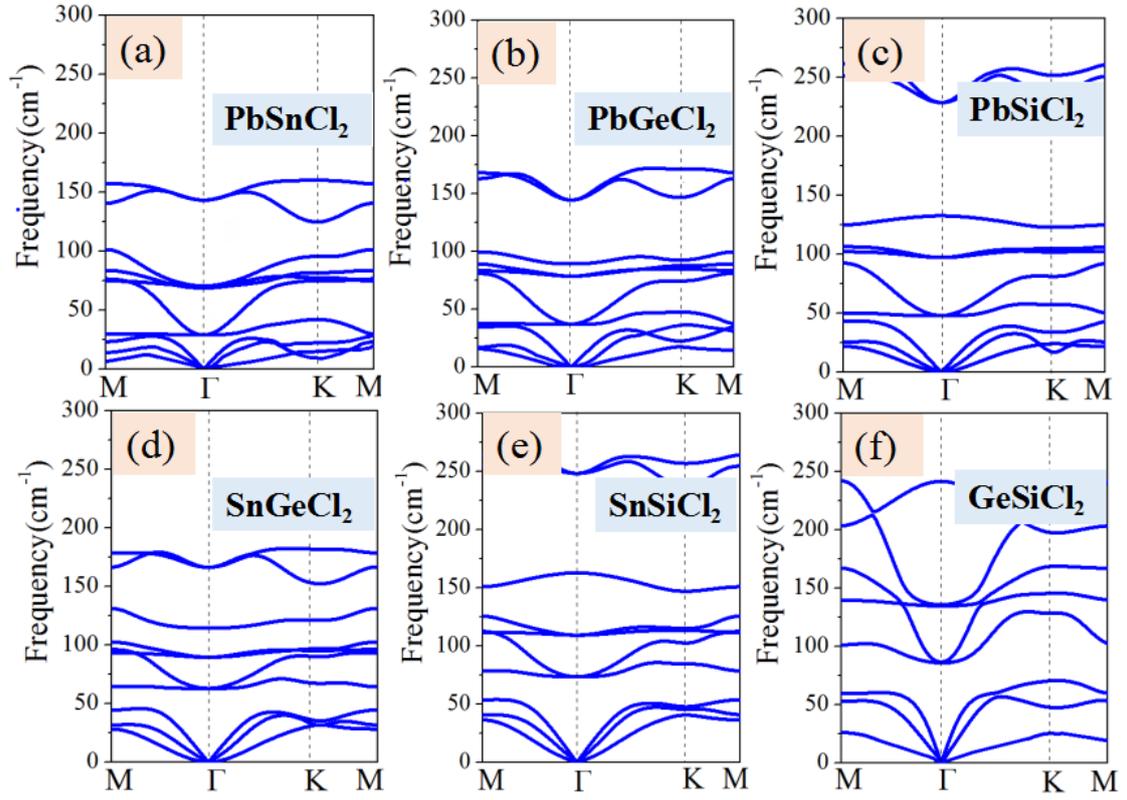

Fig. 2 Phonon band dispersions denote the PbSnCl$_2$, PbGeCl$_2$, PbSiCl$_2$, SnGeCl$_2$, SnSiCl$_2$ and GeSiCl$_2$ films.

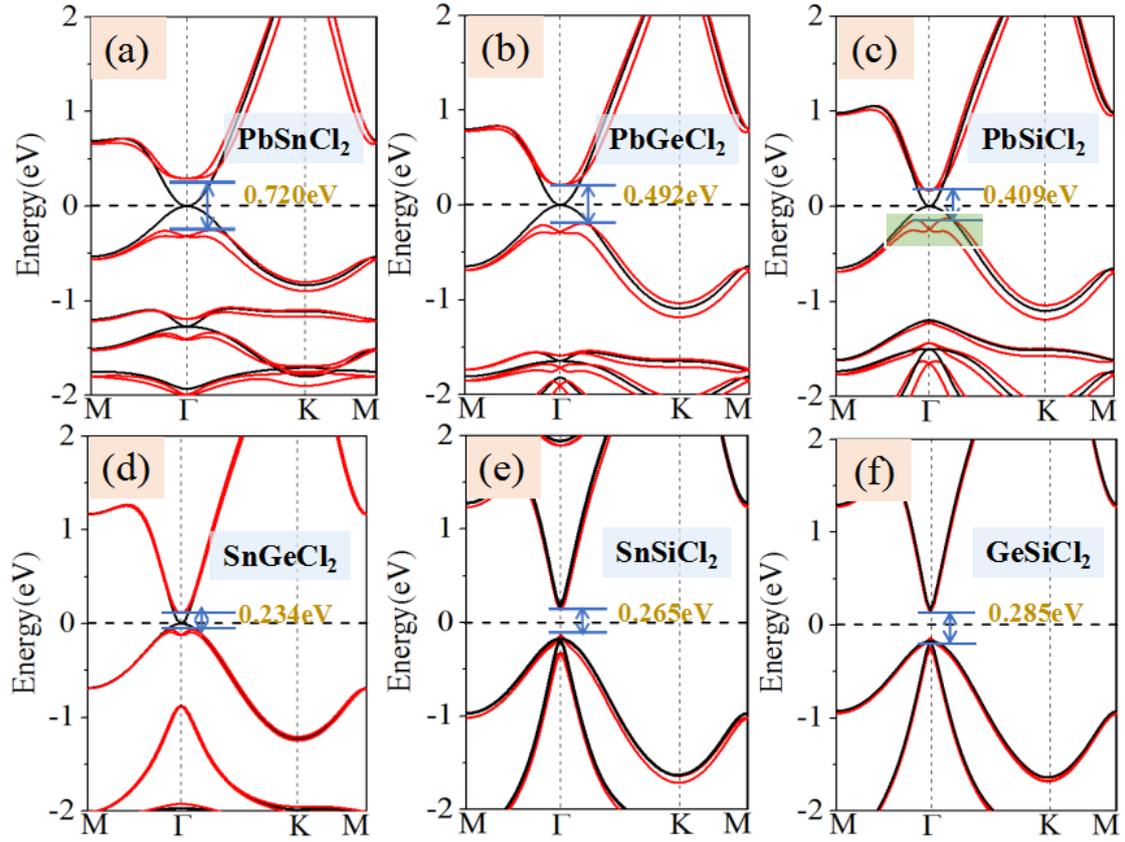

Fig. 3 The calculated band structures without and with SOC. The black lines correspond to band structures without SOC, the red lines correspond to band structures with SOC.

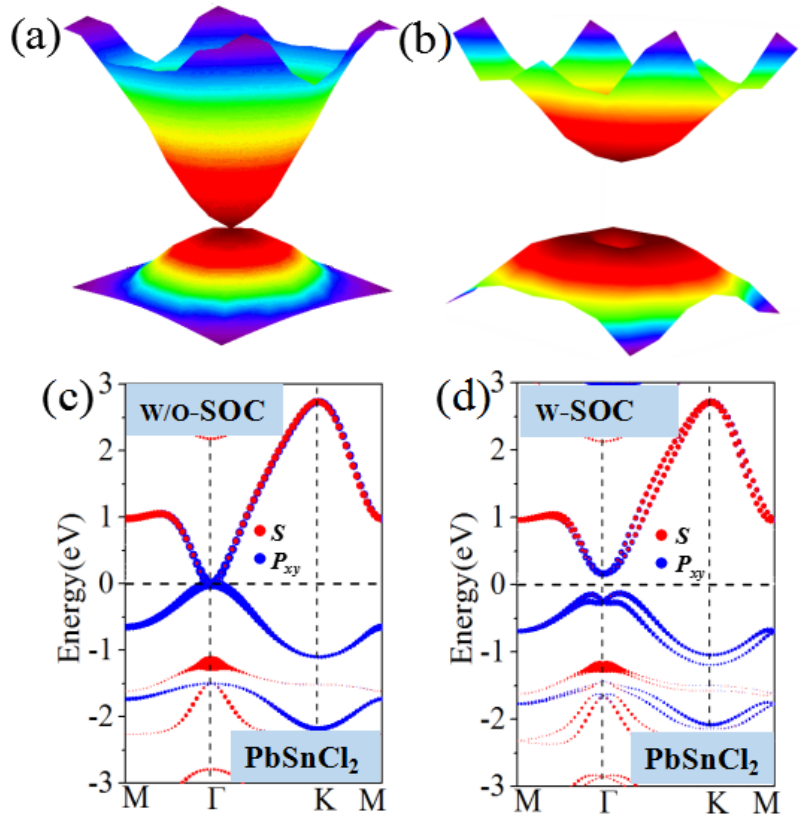

Fig. 4 (a) and (b) show 3D band structure of 2D PbSnCl$_2$ monolayer. (a) PbSnCl$_2$ monolayer without SOC. (b) PbSnCl$_2$ film with SOC. Orbitals-resolved band structures without and with SOC for (c) and (d) PbSnCl$_2$ monolayer.

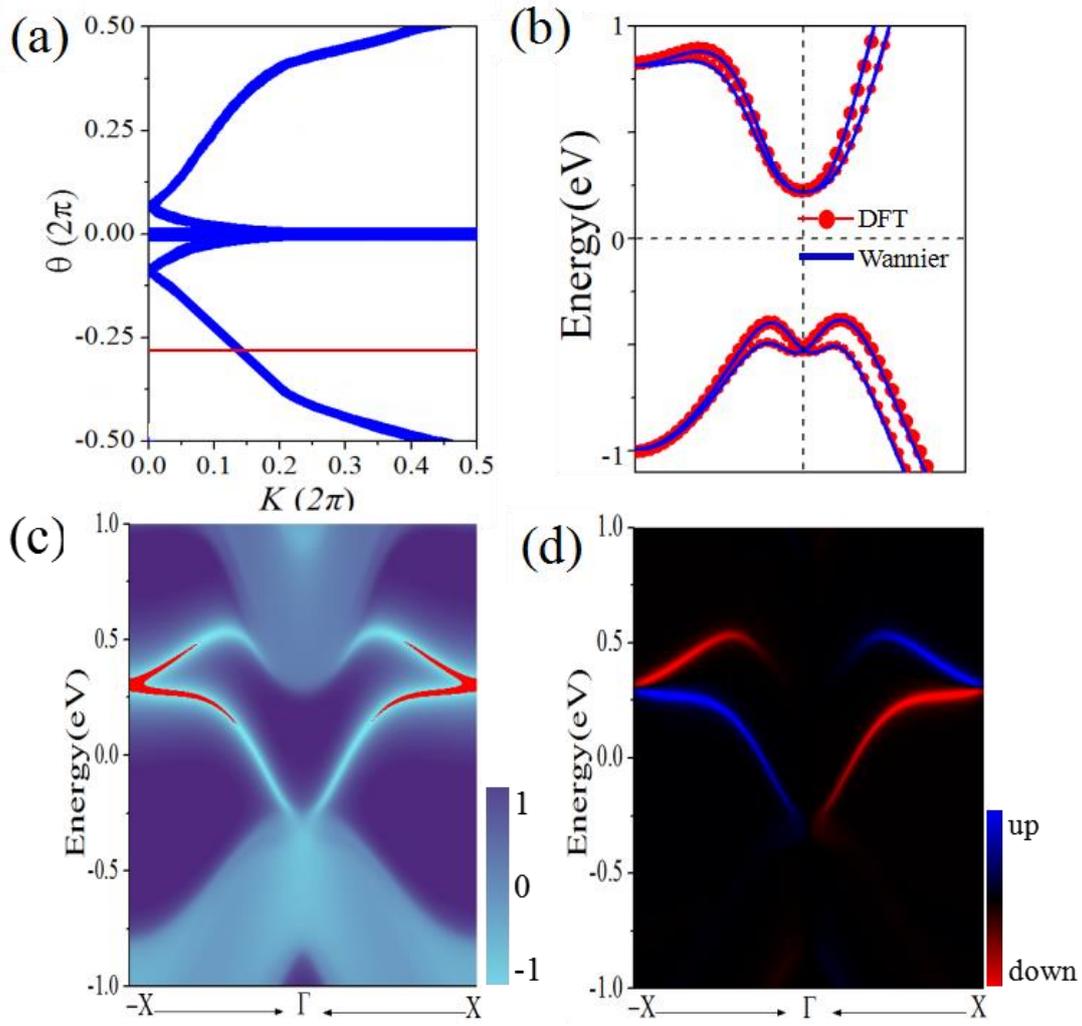

Fig. 5 (a) The evolution lines of WFC for PbSnCl$_2$ monolayer. (b) The energy bands was calculated around the Fermi level by the method of DFT and Wannier. (c) Total (left panel) and (d) spin (right panel) edge density of states for PbSnCl$_2$. In the spin edge plot, blue/red lines denote the spin up/down polarization.

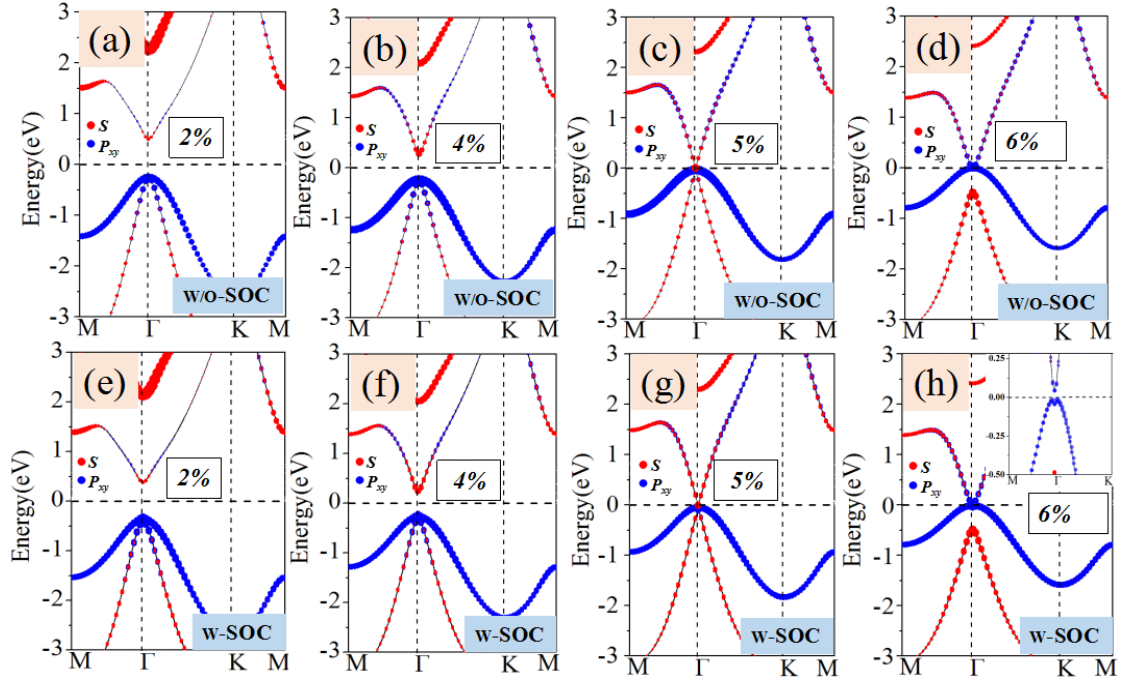

Fig. 6 The GeSiCl$_2$ energy band structures as a function of strain, the blue points shows *s* orbit composition and the red points shows $p_{x,y}$ orbit composition. (a)(b)(c) and (d) show GeSiCl$_2$ film without SOC,(e)(f)(g) and (h) show GeSiCl$_2$ film without SOC.

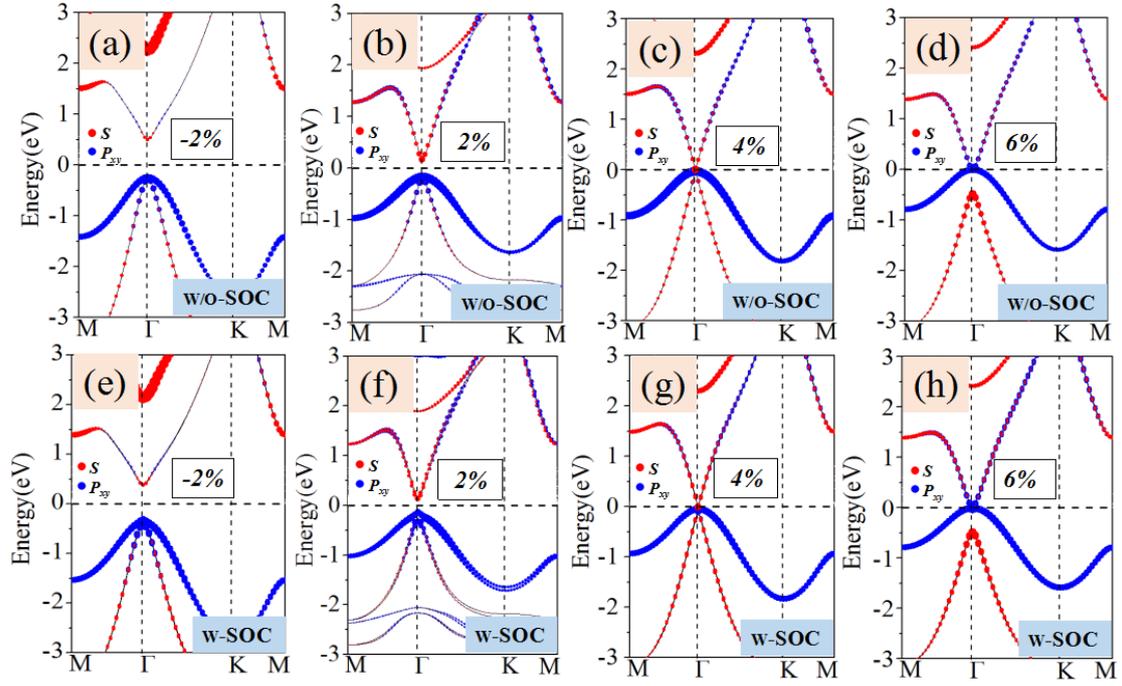

Fig. 7 The evolution of atomic *s* and $p_{x,y}$ orbital without SOC and with SOC of (a)-(d) and (e)-(h) SnSiCl$_2$ monolayer.

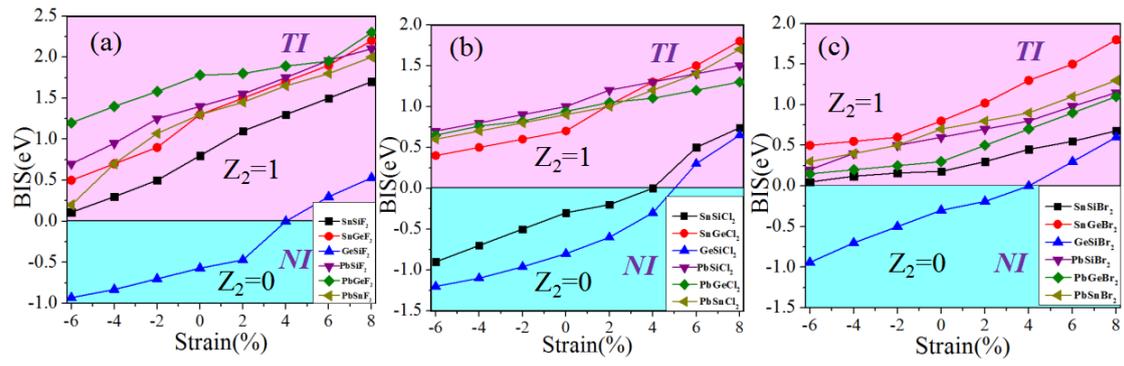

Fig. 8 Variation of band inversion strength (BIS) for $ABF_2$, $ABCl_2$ and $ABBr_2$ ($A\neq B=$ Si, Ge, Sn, Pb) systems with respect to external strain.

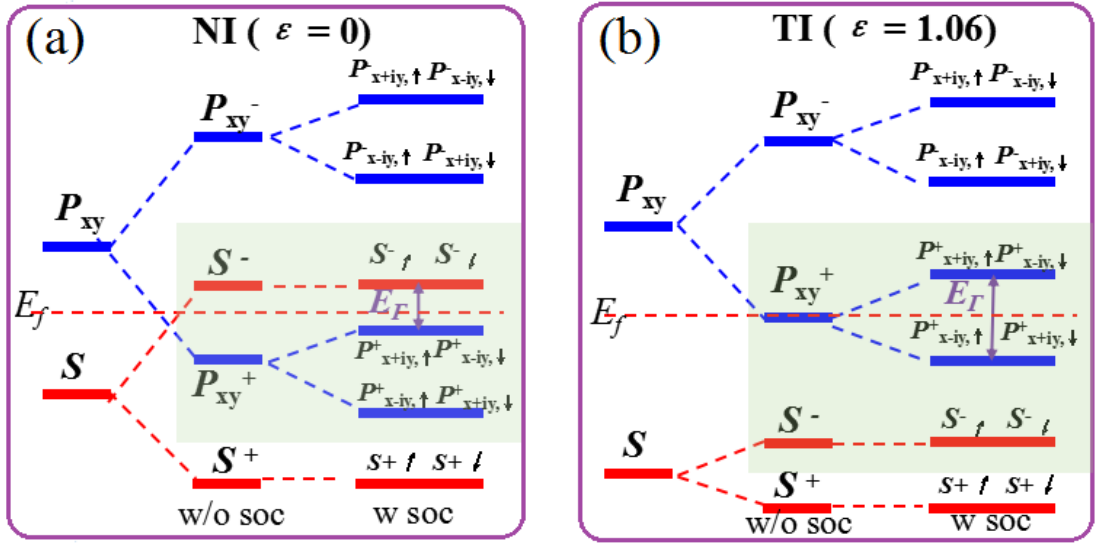

Fig. 9 The evolution of atomic *s* and $p_{x,y}$ orbital without SOC and with SOC of (a) and (b) GeSiCl$_2$. The horizontal red dashed lines indicate the Fermi level.

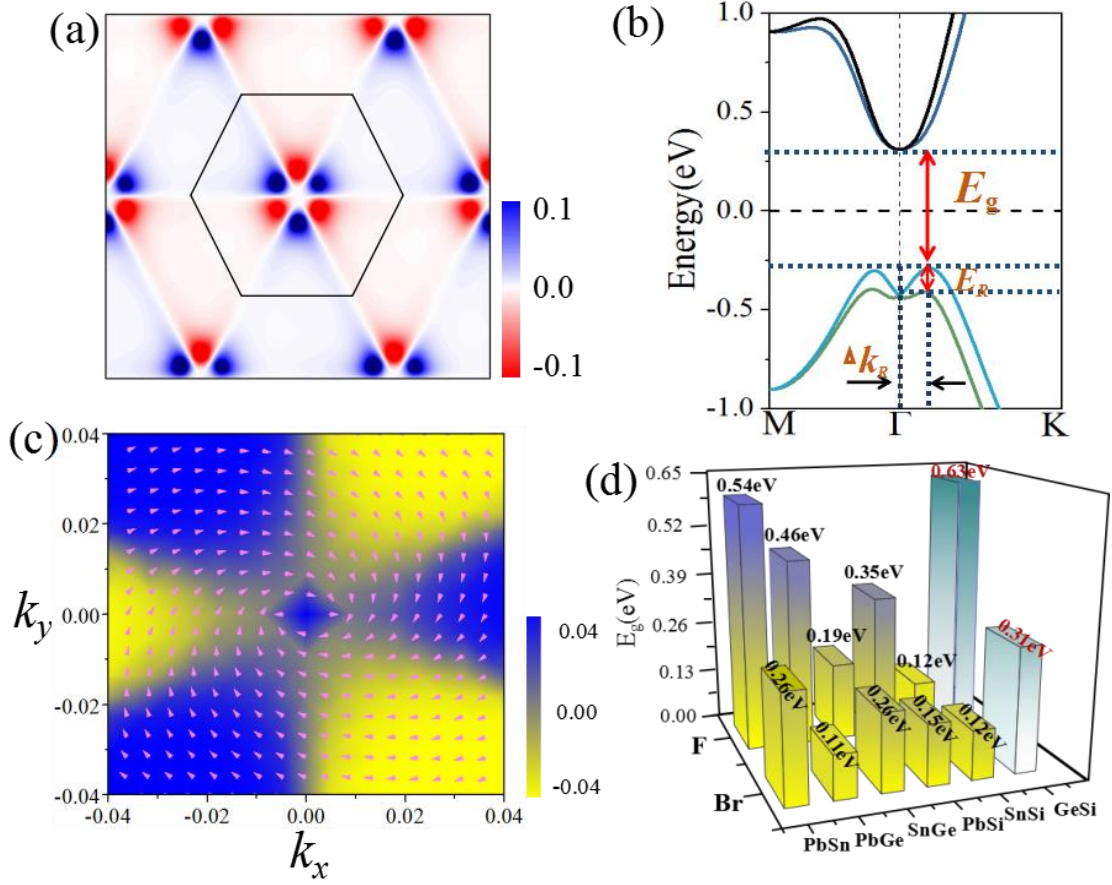

Fig. 10 (a)The k-resolved spin Berry curvature at the Fermi level of PbSnCl$_2$ monolayer. (b)The Rashba splitting energy $E_R$ and the momentum offset $k_R$ of PbSiCl$_2$ monolayer. (c)Spin texture in the highest valence bands for PbSnCl$_2$ monolayer. Arrows refer to the in-plane orientation of spin, and the color background denotes the z component of the spin. (d)The energy band for ABF$_2$, ABBr$_2$ ($A{\neq}B$= Si, Ge, Sn, Pb)systems.

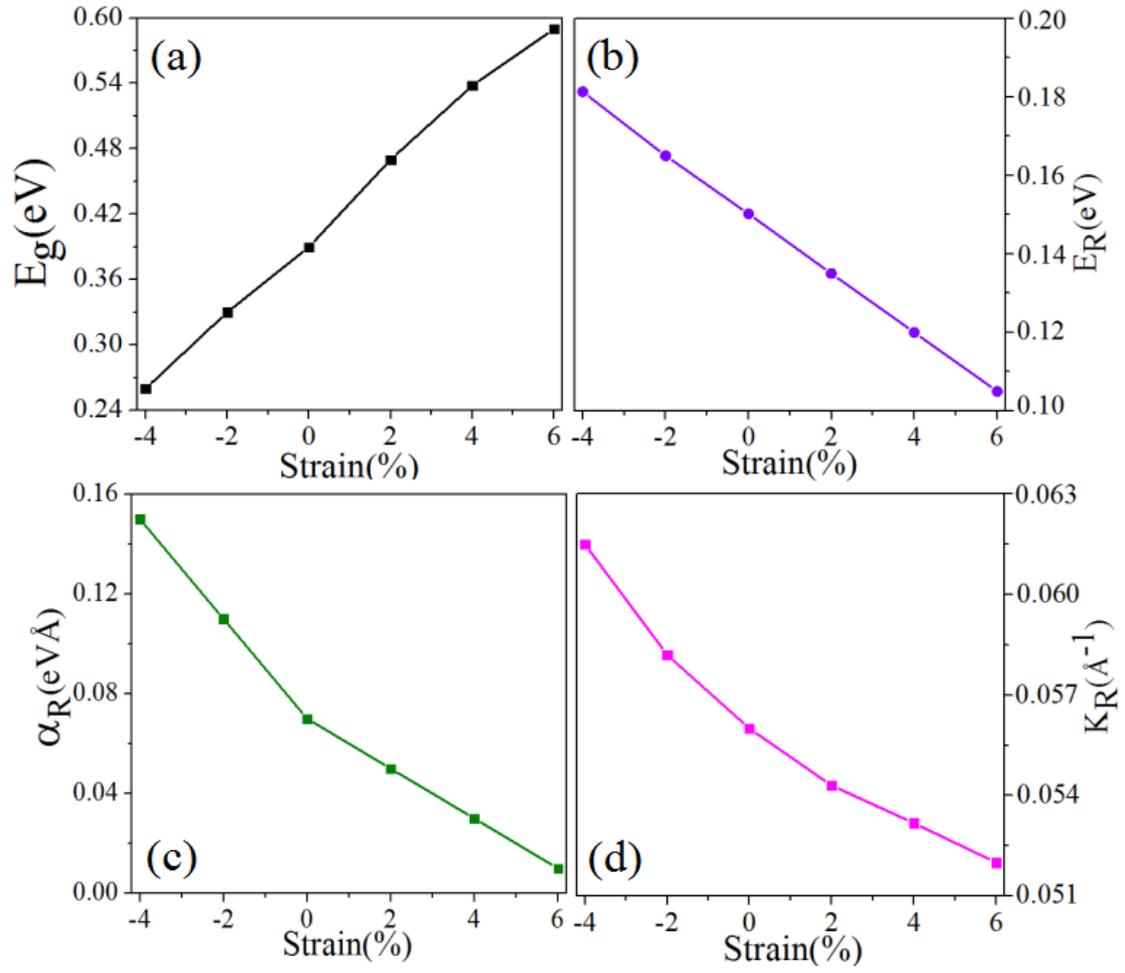

Fig. 11 Present four parameters ($E_R$, $k_R$, $α_R$, $E_g$) of PbSiCl$_2$ monolayer under different biaxial strains by solid triangles.

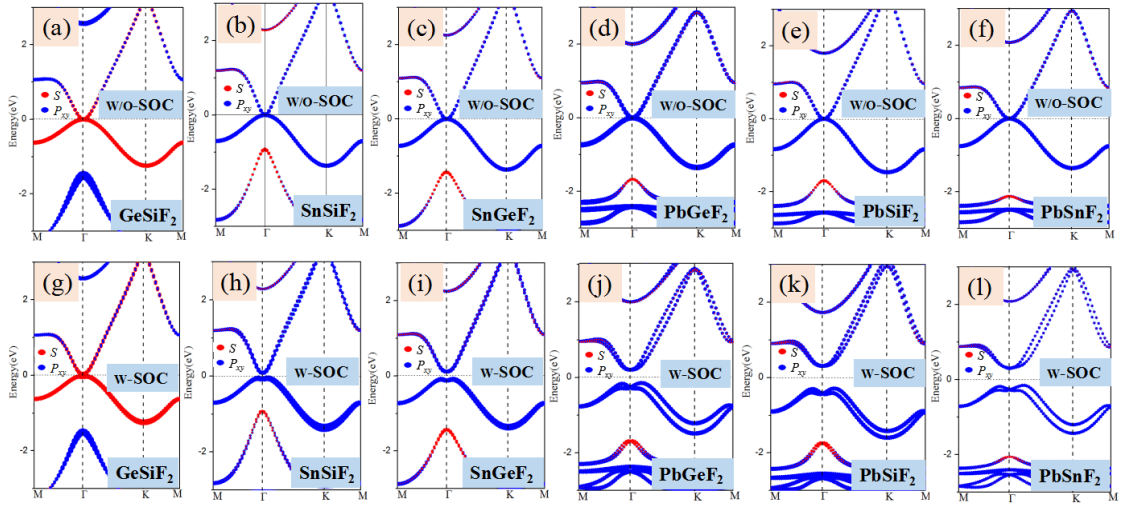

Fig. 12 The blue points shows *s* orbit composition and the red points shows $p_{x,y}$ orbit composition. (a)-(f) show ABF$_2$(A≠B= Si, Ge, Sn, Pb) films without SOC, (g)-(l) show ABF$_2$(A≠B= Si, Ge, Sn, Pb) films with SOC.

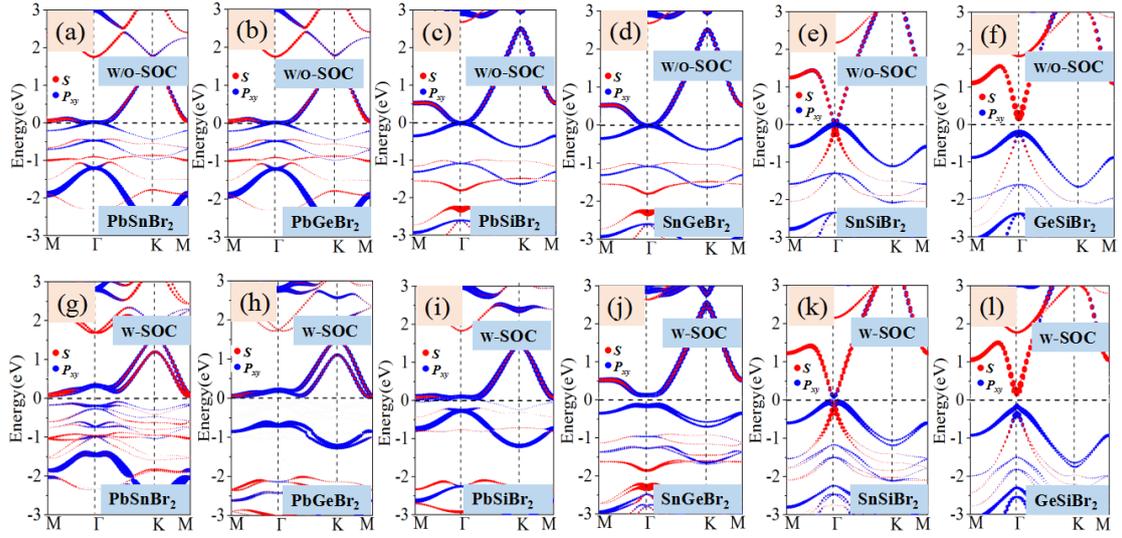

Fig. 13 The blue points shows *s* orbit composition and the red points shows $p_{x,y}$ orbit composition. (a)-(f) show ABBr$_2$($A \neq B$ = Si, Ge, Sn, Pb)films without SOC, (g)-(l) show ABF$_2$($A \neq B$= Si, Ge, Sn, Pb)films with SOC.

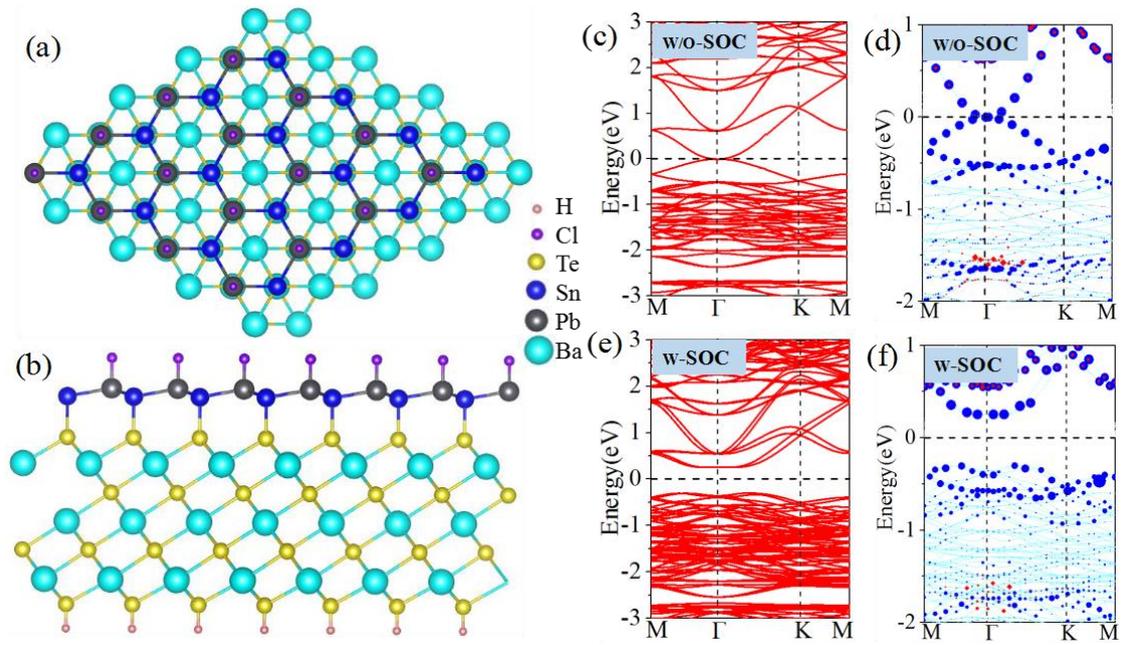

Fig. 14 Top view (a) and side view (b) of the schematic illustration of epitaxial growth PbSnCl$_2$ on Te(111)-terminated BaTe surface, as well as the band structure without SOC (c) (d) and with SOC (e) (f) for PbSnCl$_2$ monolayer.